\documentclass[prd,superscriptaddress,nofootinbib,amsmath,amssymb,aps,11pt]{revtex4-2}

\usepackage{bm}
\usepackage{amsfonts}
\usepackage{latexsym}
\usepackage[latin1]{inputenc}
\usepackage{graphicx}
\usepackage{amsmath}
\usepackage{palatino}
\usepackage{mathpazo}
\usepackage{booktabs}
\usepackage{dcolumn}

\def\jnl@style{\it}
\def\aaref@jnl#1{{\jnl@style#1}}

\def\aaref@jnl#1{{\jnl@style#1}}

\def\aj{\aaref@jnl{AJ}}                   
\def\apj{\aaref@jnl{ApJ}}                 
\def\apjl{\aaref@jnl{ApJ}}                
\def\apjs{\aaref@jnl{ApJS}}               
\def\apss{\aaref@jnl{Ap\&SS}}             
\def\aap{\aaref@jnl{A\&A}}                
\def\aapr{\aaref@jnl{A\&A~Rev.}}          
\def\aaps{\aaref@jnl{A\&AS}}              
\def\mnras{\aaref@jnl{Mon.~Not.~Roy.~Astron.~Soc.}}             
\def\prd{\aaref@jnl{Phys.~Rev.~D}}        
\def\prc{\aaref@jnl{Phys.~Rev.~C}}  
\def\prl{\aaref@jnl{Phys.~Rev.~Lett.}}    
\def\qjras{\aaref@jnl{QJRAS}}             
\def\skytel{\aaref@jnl{S\&T}}             
\def\ssr{\aaref@jnl{Space~Sci.~Rev.}}     
\def\zap{\aaref@jnl{ZAp}}                 
\def\nat{\aaref@jnl{Nature}}              
\def\aplett{\aaref@jnl{Astrophys.~Lett.}} 
\def\apspr{\aaref@jnl{Astrophys.~Space~Phys.~Res.}} 
\def\physrep{\aaref@jnl{Phys.~Rep.}}      
\def\physscr{\aaref@jnl{Phys.~Scr}}       
\def\commat{\aaref@jnl{Comm.~Math.~Phys.}}              
\def\science{\aaref@jnl{Science}}               
\def\cqg{\aaref@jnl{Classical Quant.~Grav.}}            
\def\jpcs{\aaref@jnl{JPCS}}                                     
\def\ijmpd{\aaref@jnl{Int.~J.~Mod.~Phys.~D}}                    
\def\grg{\aaref@jnl{Gen.~Relat.~Gravit.}}               
\def\rpp{\aaref@jnl{Rep.~Prog.~Phys.}}          
\def\npa{\aaref@jnl{Nucl.~Phys.~A}}        
\def\lrr{\aaref@jnl{Living Rev.~Rel.}}                   
\def\jcap{\aaref@jnl{J.~Cosmology Astropart.~Phys.}}    
\def\rmp{\aaref@jnl{Rev.~Mod.~Phys.}}   

\allowdisplaybreaks[1]

\begin{document}

	\title{On the dynamics of the nonrotating and rotating black hole scalarization}
	
	\author{Daniela D. Doneva}
	\email{daniela.doneva@uni-tuebingen.de}
	\affiliation{Theoretical Astrophysics, Eberhard Karls University of T\"ubingen, T\"ubingen 72076, Germany}

	\author{Stoytcho S. Yazadjiev}
	\email{yazad@phys.uni-sofia.bg}
	\affiliation{Theoretical Astrophysics, Eberhard Karls University of T\"ubingen, T\"ubingen 72076, Germany}
	\affiliation{Department of Theoretical Physics, Faculty of Physics, Sofia University, Sofia 1164, Bulgaria}
	\affiliation{Institute of Mathematics and Informatics, 	Bulgarian Academy of Sciences, 	Acad. G. Bonchev St. 8, Sofia 1113, Bulgaria}


	\begin{abstract}
Even though black hole scalarization is extensively studied recently, little has been done in the direction of understanding the dynamics of this process, especially in the rapidly rotating regime. In the present paper, we focus exactly on this problem by considering the nonlinear dynamics of the scalar field while neglecting the backreaction on the spacetime metric. This approach has proven to give good results in various scenarios and we have explicitly demonstrated its accuracy for nonrotating black holes especially close to the bifurcation point.  We have followed the evolution of a black hole from a small initial perturbation, throughout the exponential growth of the scalar field followed by a subsequent saturation to an equilibrium configuration. As expected, even though the emitted signal and the time required to scalarize the black hole are dependent on the initial perturbation, the final stationary state that is reached is independent on the initial data.        
	\end{abstract}
	
	\maketitle
	
	\section{Introduction}
In extended scalar-tensor theories, such  as Gauss-Bonnet gravity, the black holes can undergo spontaneous scalarization -- 
a strong gravity phase transition triggered by a tachyonic instability  due to the non-minimal coupling between the scalar field(s) and the spacetime curvature \cite{Doneva:2017bvd,Silva:2017uqg,Antoniou:2017acq}.  This very interesting phenomenon is the only known dynamical mechanism for endowing black holes (and other compact
objects) with scalar hair without altering the predictions in the weak field limit. The tachyonic instability that triggers the spontaneous
scalarization is to a large extent well understood and extensively studied both in the static and the rotating regimes \cite{Doneva2010,Gao:2018acg,Dima:2020yac,Hod:2020jjy,Doneva:2020nbb,Doneva:2020kfv,Macedo:2020tbm,Zhang:2020pko}. The same applies to the scalarized solutions produced in this way, both in the static and rotating cases \cite{Doneva:2017bvd,Silva:2017uqg,Cunha:2019dwb,Collodel:2019kkx,Herdeiro:2020wei,Berti:2020kgk}, as well as their stability \cite{Blazquez-Salcedo:2018jnn,Blazquez-Salcedo:2018tyn,Blazquez-Salcedo:2018tyn,Minamitsuji:2018xde,Silva:2018qhn}. Black hole scalarization was also addressed in other more general extended  scalar-tensor theories \cite{Ventagli:2020rnx,Andreou:2019ikc}. The very dynamics of the spontaneous scalarization, i.e. the process of forming scalarized black holes from non-scalarized ones, is much less studied and understood. There are only few exceptions. A toy model of the  dynamics of spontaneous scalarization for spherically symmetric configurations within Maxwell-scalar models in flat spacetime was recently studied in \cite{Herdeiro:2020htm}. 
A  fully non-linear numerical evolution of the scalarization in the spherically symmetric case within the Einstein-Maxwell-scalar gravity was performed in \cite{Herdeiro:2018wub}. The dynamical scalarization and descalarization in binary black hole mergers was examined in the decoupling limit in \cite{Silva:2020omi}.  Different aspects of the scalarization dynamics of anther type of compact objects, the neutron stars, were considered as well in a number of papers \cite{Novak1998,Novak1998a,Barausse2013,Shibata2014,Gerosa:2016fri,Sperhake:2017itk}.

However, in the  case of realistic rotating single black holes there is no study of the dynamics of the spontaneous scalarization.  The purpose of the present paper is to make a first step in this direction by considering static and rotating isolated black holes in scalar-Gauss-Bonnet gravity. Solving the scalarization dynamics in its full generality and nonlinearity for rotating black holes is a formidable task and that is why in the present paper we consider a nonlinear but simplified model. We study the spontaneous scalarization dynamics in the ``decoupling limit'' -- we numerically evolve the nonlinear scalar field equation on the fixed geometric background
of a Kerr black hole (or a Schwarzschild black hole as a particular case). In other words we neglect the back reaction of the scalar field dynamics on the spacetime geometry. Although simplified we believe that the model captures the basic qualitative features of the full nonlinear dynamics and can  give a good insight into the general case. Moreover, as we will see later, in the case of static black  holes and for certain coupling functions  our approach leads to results which are in a very good agreement with the solutions obtained when solving the nonlinear static field equations. In any case, considering the decoupling limit is a very good approximation of the full nonlinear picture  in the vicinity of the bifurcation point where the back reaction of the scalar field on the geometry is small.         

In Section II we discuss the basic formulation of the problem and the approach we are adopting. This is followed by  Section III, where first the numerical code is briefly discussed and afterwards the results are presented for the static and the rotating black holes. For nonotating solution not only the evolution of the scalar field is discussed, but also the accuracy of the approximation we employed is evaluated. The paper ends with Conclusions.

	\section{General equations and setting the problem}	
	The scalar-Gauss-Bonnet gravity in vacuum  is defined by the action   
	
	\begin{eqnarray}
		S=&&\frac{1}{16\pi}\int d^4x \sqrt{-g} 
		\Big[R - 2\nabla_\mu \varphi \nabla^\mu \varphi 
		+ \lambda^2 f(\varphi){\cal R}^2_{GB} \Big] ,\label{eq:quadratic}
	\end{eqnarray}
	where $R$ is the Ricci scalar with respect to the spacetime metric $g_{\mu\nu}$, $\varphi$ denotes the scalar field  with a coupling function  $f(\varphi)$, $\lambda$ is the so-called Gauss-Bonnet coupling constant having  dimension of $length$ and ${\cal R}^2_{GB}=R^2 - 4 R_{\mu\nu} R^{\mu\nu} + R_{\mu\nu\alpha\beta}R^{\mu\nu\alpha\beta}$ is the Gauss-Bonnet invariant with $R_{\mu\nu\alpha\beta}$ being the 
	curvature tensor. The variation of this action with respect to the metric $g_{\mu\nu}$ and scalar field $\varphi$ gives the following field equations 
			
	\begin{eqnarray}\label{FE}
		&&R_{\mu\nu}- \frac{1}{2}R g_{\mu\nu} + \Gamma_{\mu\nu}= 2\nabla_\mu\varphi\nabla_\nu\varphi -  g_{\mu\nu} \nabla_\alpha\varphi \nabla^\alpha\varphi ,\\
		&&\nabla_\alpha\nabla^\alpha\varphi=  -  \frac{\lambda^2}{4} \frac{df(\varphi)}{d\varphi} {\cal R}^2_{GB},\label{SFE}
	\end{eqnarray}
	where  $\nabla_{\mu}$ is the covariant derivative with respect to  $g_{\mu\nu}$ and  $\Gamma_{\mu\nu}$ is defined by 
	\begin{eqnarray}
		&&\Gamma_{\mu\nu}= - R(\nabla_\mu\Psi_{\nu} + \nabla_\nu\Psi_{\mu} ) - 4\nabla^\alpha\Psi_{\alpha}\left(R_{\mu\nu} - \frac{1}{2}R g_{\mu\nu}\right) + 
		4R_{\mu\alpha}\nabla^\alpha\Psi_{\nu} + 4R_{\nu\alpha}\nabla^\alpha\Psi_{\mu} \nonumber \\ 
		&& - 4 g_{\mu\nu} R^{\alpha\beta}\nabla_\alpha\Psi_{\beta} 
		+ \,  4 R^{\beta}_{\;\mu\alpha\nu}\nabla^\alpha\Psi_{\beta} 
	\end{eqnarray}  
	with 
	\begin{eqnarray}
		\Psi_{\mu}= \lambda^2 \frac{df(\varphi)}{d\varphi}\nabla_\mu\varphi .
	\end{eqnarray}

	In the present paper  we will consider  asymptotically flat spacetimes and  the case for which the cosmological value of the scalar field is zero, i.e. $\varphi_{\infty}=0$. The  Gauss-Bonnet coupling function $f(\varphi)$ allowing for spontaneous scalarization has to obey the conditions $f(0)=0$, $\frac{df}{d\varphi}(0)=0$  and $\frac{d^2f}{d\varphi^2}(0)=\epsilon$ with $\epsilon=\pm 1$. With these conditions
	imposed on the coupling function  the Kerr black hole solution is also 
	a black hole solution to the scalar-Gauss-Bonnet  gravity with a trivial scalar field  $\varphi=0$. However, within a certain region of the parameter space, the Kerr solution is unstable 
	 within the bigger theory against linear scalar perturbations -- it suffers from a tachyonic instability.  The general expectation
	 is that the exponential growth of the scalar field will last until the scalar field becomes large enough so that the nonlinear terms in
	 the coupling function suppress the instability giving rise to a new stationary scalarized solution. 
	 
	 Our purpose in this work is 
	 to show that the described qualitative pictures really happens. As we have already mentioned, the fully nonlinear dynamical problem is extremely difficult, especially in the case of rotating black holes. That is why we shall base our study on a simplified model which is free from heavy technical complications but preserves the leading role of the nonlinearity associated with the coupling function. We shall consider a model where the spacetime geometry is kept fixed and 
	 the whole dynamics is governed by the nonlinear equation for the scalar field. This dynamical model is a very good approximation for example in the vicinity of the bifurcation point where the back reaction of the scalar field on the geometry is small. In addition, similar type of approximations were successfully applied in the studies of binary black hole mergers \cite{Witek:2018dmd,Silva:2020omi}.
	 Therefore, we will consider the nonlinear wave equation  (\ref{SFE}) on the Kerr background geometry  which, in the standard Boyer-Lindquist coordinates, reads
	 
	\begin{eqnarray}\label{KerrM}
		ds^2= &&- \frac{\Delta -a^2\sin^2\theta}{\Sigma} dt^2 - 2a \sin^2\theta \frac{r^2 + a^2 - \Delta}{\Sigma} dt d\phi   \\
		&&+ \frac{(r^2 + a^2)^2 - \Delta a^2 \sin^2\theta}{\Sigma} \sin^2\theta d\phi^2 + \frac{\Sigma}{\Delta} dr^2 + \Sigma d\theta^2 \nonumber
	\end{eqnarray} 
	where $\Delta=r^2 - 2Mr + a^2$ and $\Sigma=r^2 + a^2 \cos^2\theta$. The Gauss-Bonnet invariant for the Kerr geometry is explicitly given by 
	\begin{eqnarray}
		{\cal R}^2_{GB}= \frac{48 M^2}{\Sigma^6}(r^2- a^2\cos^2\theta)(r^4 - 14 a^2 r^2 \cos^2\theta + a^4\cos^4\theta).
	\end{eqnarray} 
	
	Before writing the explicit form of the perturbation equation (\ref{SFE}) we introduce a new azimuthal coordinate $\phi_*$ defined by 
	\begin{eqnarray}
		d\phi_*=d\phi + \frac{a}{\Delta} dr.  
	\end{eqnarray} 
	Working with $\phi_*$ allows us to get rid of some unphysical pathologies near the horizon. It is also convenient to work with the tortoise coordinate $x$ defined by 
	
	\begin{eqnarray}
		dx= \frac{r^2+a^2}{\Delta} dr.  
	\end{eqnarray} 
	
	In the coordinates $(t,x,\theta,\phi_*)$ the  equation (\ref{SFE}) takes the following explicit form on the Kerr background
 	
	\begin{eqnarray} \label{eq:PertEq}
		&&-\left[(r^2 + a^2)^2 - \Delta a^2 \sin^2\theta\right] \partial^2_t \varphi + (r^2 + a^2)^2 \partial^2_x \varphi + 2r \Delta \partial_x\delta\varphi - 4Ma r\partial_t\partial_{\phi_*}\varphi \nonumber \\ 
		&&+  2a(r^2 + a^2)\partial_x\partial_{\phi_*}\varphi  + \Delta\left[\frac{1}{\sin\theta} \partial_\theta(\sin\theta\partial_\theta\varphi) +   \frac{1}{\sin^2\theta}\partial^2_{\phi_*}\varphi \right] \\ 
		&& = - \lambda^2 \frac{12 M^2\Delta}{\Sigma^5}(r^2- a^2\cos^2\theta)(r^4 - 14 a^2 r^2 \cos^2\theta + a^4\cos^4\theta)\frac{df(\varphi)}{d\varphi}. \nonumber
	\end{eqnarray}

	The boundary conditions we have to impose when evolving in time eq. \eqref{eq:PertEq} is that the scalar field perturbation has the form of an outgoing wave at infinity and an ingoing wave at the black hole horizon. From  a numerical point of view the boundary condition at infinity is more difficult to be implemented and any inaccuracies inevitably lead to undesidered  reflected  wave from infinity. 
One possibility to prevent the reflected wave from spoiling the extracted signal is to push the outer boundary of the numerical grid sufficiently far away, such that the spurious reflection returns to the location at which the signal is extracted only after a time such that the unspoiled signal has a long enough duration for a satisfactorily precise analysis.

		\section{Results}

		We will consider two coupling functions
		\begin{eqnarray} 
		&&f(\varphi)_{\rm Case I} = \frac{\epsilon}{2\beta} \left(1 - e^{-\beta \varphi^2}\right) , \label{eq:coupling12} \\
		&&f(\varphi)_{\rm Case II} = \frac{\epsilon}{\beta^2}\left(1 - \frac{1}{\cosh(\beta \varphi)}\right) , \label{eq:coupling21}
		\end{eqnarray} 
		where $\beta$ is a parameter and $\epsilon=\pm 1$. For $\epsilon= +1$ both rotating and nonrotating black holes can scalarized while the case of $\epsilon=-1$ is responsible for the spin-induced scalarization where black holes can scalarize only if they are rotating fast enough. The behavior of the static scalarized black hole solutions for different values of $\beta$ was considered in detail in \cite{Doneva:2018rou} and it can be summarized in the following way. It turns out that both \eqref{eq:coupling12} and \eqref{eq:coupling21} lead to well behaved scalarized black hole branches of solutions that are stable (at least away from the region where a loss of  the hyperbolic character of the field equations is observed \cite{Blazquez-Salcedo:2018jnn}). The general observation, at least for the range of parameters considered in \cite{Doneva:2018rou}, is that the increase of $\beta$ can increase the range of masses where scalarized black holes exist. In addition, larger $\beta$ suppress the scalar field and thus lead to smaller deviations from general relativity. This is a particularly desired behavior in our approximate nonlinear treatment of the scalarization, since smaller differences from the Kerr black hole will most probably lead to a better accuracy of the decoupling limit approximation on the overall dynamics of the system. We have performed the simulations for two values of $\beta$, namely $\beta=6$ and $\beta=12$. While the first case is the one considered in the original paper on black hole scalarization \cite{Doneva:2017bvd} and also for the rotating solutions in \cite{Cunha:2019dwb,Herdeiro:2020wei}, the second one leads to almost perfect agreement between our approximate approach and the fully nonlinear static black hole solutions as shown below. The qualitative behavior in both cases is quite similar, though, and in order to keep the presentation more tractable the results in the following sections will be presented only for $\beta=12$ where the deviations from the true nonlinear results are expected to be much smaller.
		 
	The numerical codes for the time evolution are based on the one developed in \cite{Blazquez-Salcedo:2018jnn} for the nonrotating case and \cite{Doneva:2020nbb} for the rotating case, with the necessary adjustments to handle the nonlinearity in the scalar field appearing on the right-hand side of equation \eqref{eq:PertEq}. The two versions of the code for $a=0$ and $a>0$ were made just for convenience since in the nonrotating case we can deal only with one spacial dimension. Thus equation \eqref{eq:PertEq} is significantly simplified and the simulations are considerably faster. This was desired because we have performed a very large number of simulations for $a=0$ in order to check the accuracy of the decoupling limit approximation for different $f(\varphi)$ and $\beta$. The two codes are in a perfect agreement in the $a=0$ limit. 
	
	We have performed a number of tests that demonstrate the correctness of the developed codes:
	\begin{itemize}
		\item In the case when the right-hand side of eq. \eqref{eq:PertEq} is neglected, the quasinormal mode frequencies of the Kerr  black hole are reproduced.
		\item For $\beta=0$, i.e. when the nonlinearities are practically neglected for the considered coupling functions, one can obtain the position of the bifurcation points for different $M/\lambda$ and $a/\lambda$ similar to \cite{ Doneva:2018rou,Doneva:2020nbb} \footnote{Note that only the derivative of the coupling function enters the scalar field equation \eqref{eq:PertEq} and that is why taking the limit $\beta \rightarrow 0$ is trivial and non-problematic.}.  The results coincide with the ones available in the literature both for static and rotating black holes \cite{Doneva:2017bvd,Cunha:2019dwb,Dima:2020yac,Doneva:2020nbb,Herdeiro:2020wei}.
		\item As we will discuss in detail below, in the region where the Schwarzschild black hole is unstable, the resulting scalar field profile at the end of the evolution is very similar to the one obtained from the solution of the static dimensionally reduced nonlinear field equations \cite{Doneva:2017bvd}.   
	\end{itemize} 
	
	 Below we will focus separately on the cases of static and rotating black holes, including the case of spin-induced scalarization. All dimensional quantities will be normalized with respect to $\lambda$.
		
	\subsection{Nontoratating black holes}
	If one performs time evolution of unstable Schwarzschild black hole within Gauss-Bonnet theory, the scalarization manifests itself as an exponential growth of the scalar field, independent of the initial conditions. In the linearized case considered in \cite{Doneva:2020nbb} the scalar field will naturally increase indefinitely. This exponential growth is halted if the full form of the scalar field equation is considered, including as well the nonlinearities. In this way we are taking into account the backreaction of the  scalar field. Of course, considering only the evolution of the scalar field without evolving the metric moves us away from the complete description of the scalarization dynamics. It is natural to expect, though, that close to the bifurcation point, where the Schwarzschild solution loses stability, this approximation is well satisfied. We will demonstrate this below and even more -- it turns out that for some of the coupling functions this approximation is satisfied with a very good accuracy even far away from the bifurcation.  Thus the purpose of this subsection is twofold -- first, to study the dynamics of the development of a scalar field and second, to  study up to what extend the approximate approach is justified since this is much easier to be evaluated in the case of spherical symmetry. The experience obtained here will be transferred to the next section where rotating solutions are considered.

	\begin{figure}
		\includegraphics[width=1\textwidth]{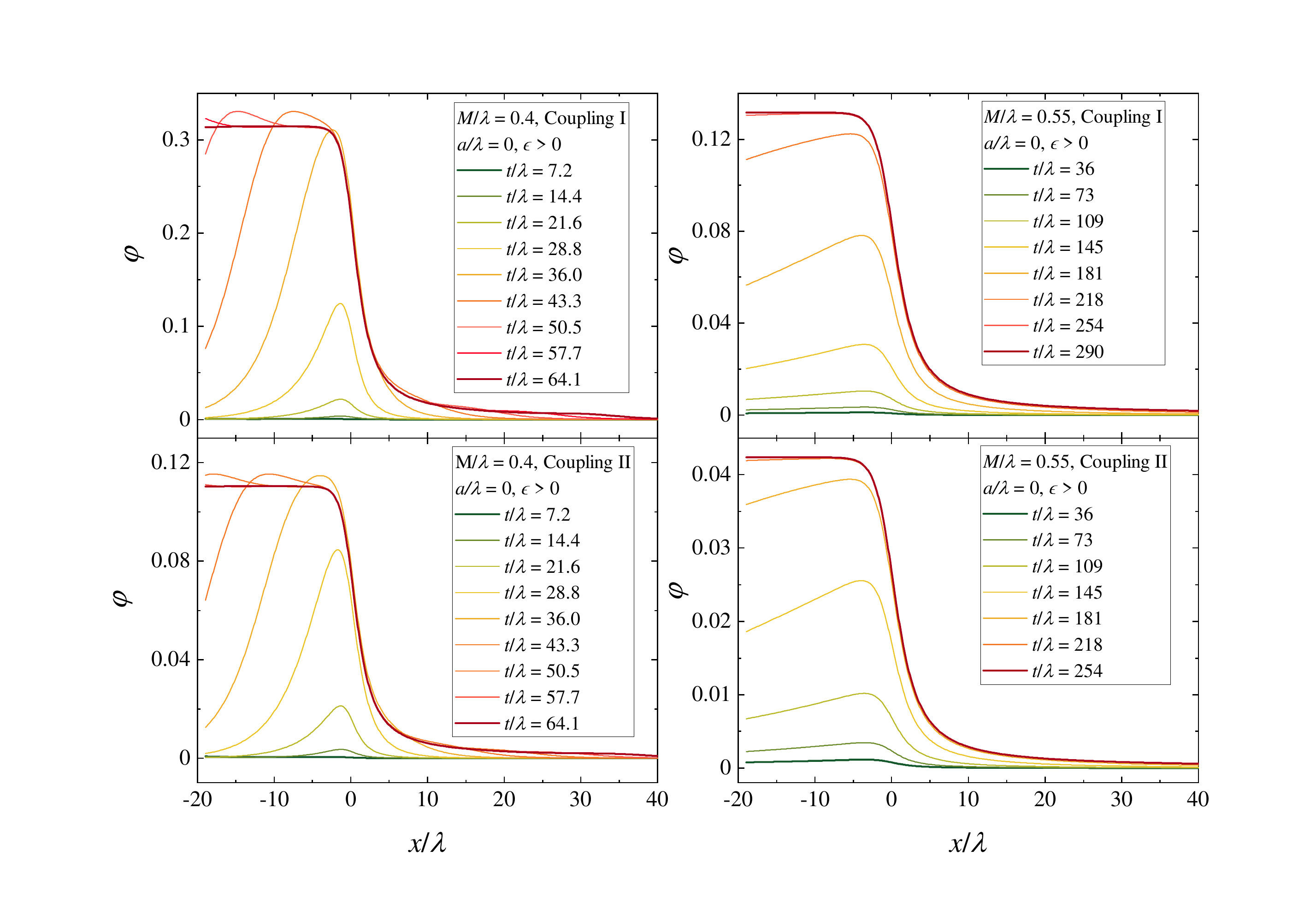}
		\caption{Time evolution of the scalar field radial profile for several representative models (presented in the normalized tortoise coordinate $x/\lambda$ on the $x$-axis) for the two coupling functions \eqref{eq:coupling12} \textit{(top panels)} and \eqref{eq:coupling21} \textit{(bottom panels)}, where $\beta=12$, and for two black hole masses $M/\lambda=0.4$\textit{ (left panels) }and $M/\lambda=0.55$ \textit{(right panels)}. Different colors of the lines represent the scalar field radial profile at different time steps. The solid dark red line represents the final state that is achieved. }
		\label{fig:Psi_evol_nonrot}
	\end{figure}
	
	\begin{figure}
		\includegraphics[width=0.45\textwidth]{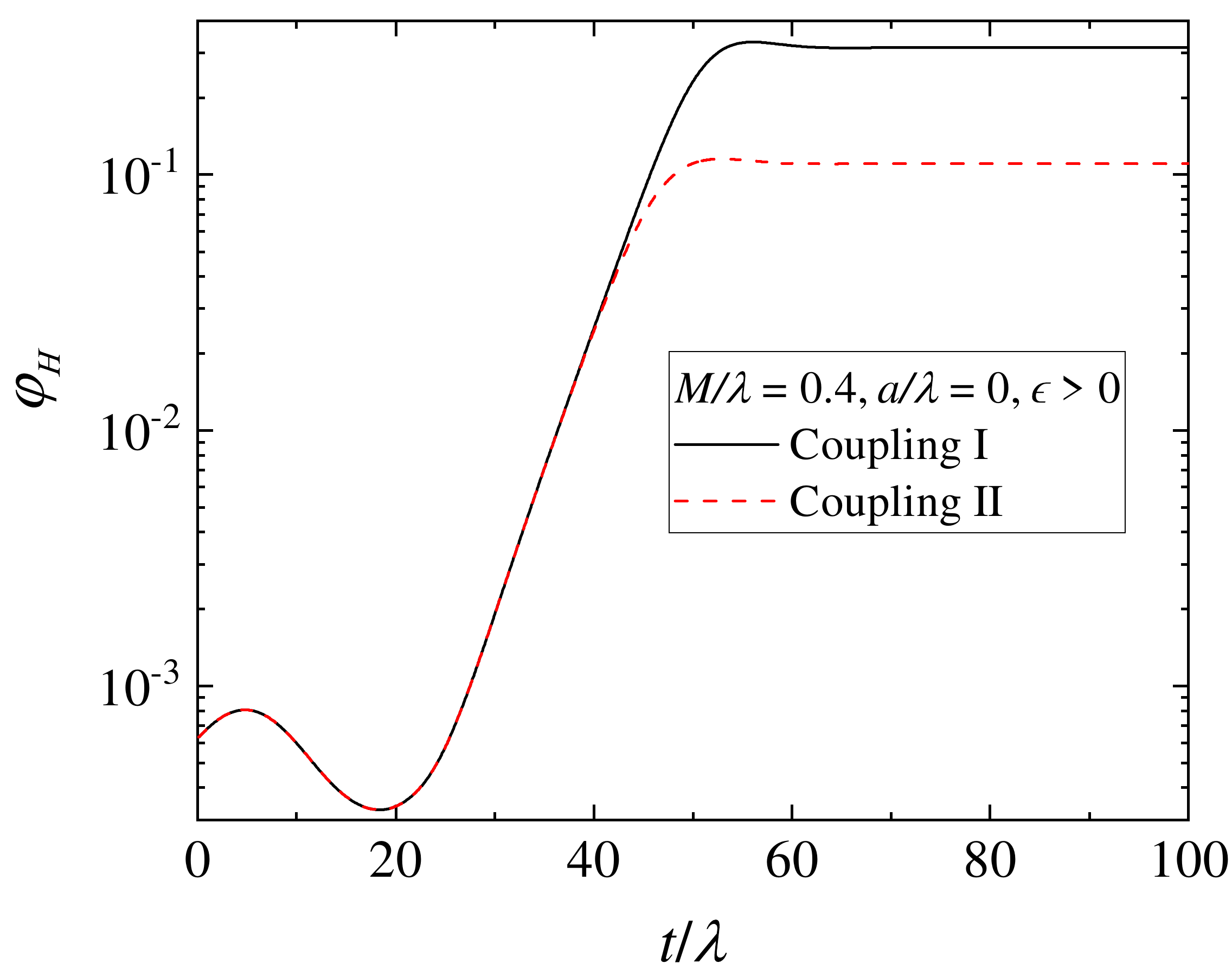}
		\includegraphics[width=0.45\textwidth]{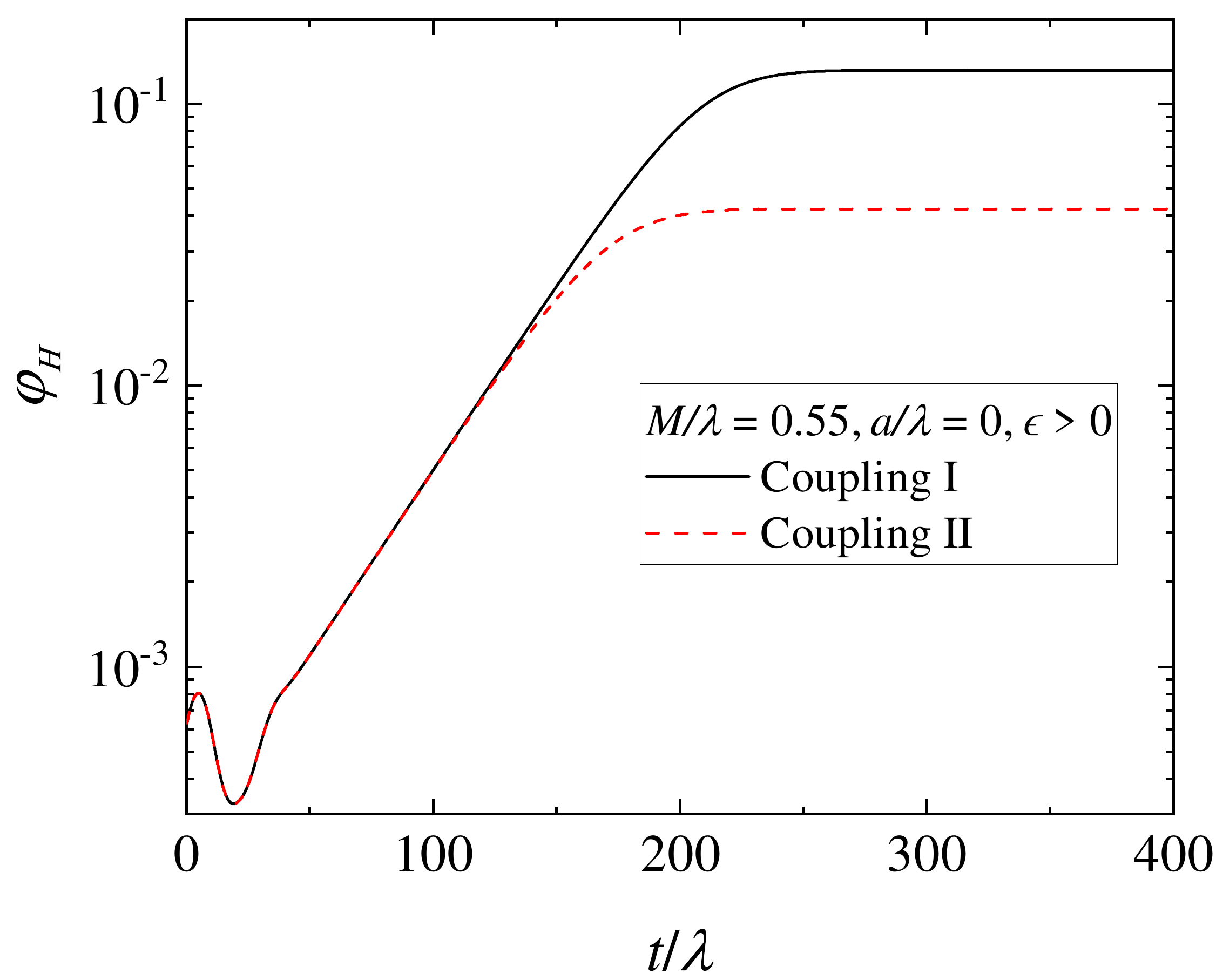}
		\caption{The  scalar field at the horizon as a function of time, for the same models presented in Fig. \ref{fig:Psi_evol_nonrot}. }
		\label{fig:phiH_t_nonrot}
	\end{figure}

	 Snapshots of the scalar field radial profile at different times are depicted in Fig. \ref{fig:Psi_evol_nonrot} while the time evolution of the scalar field at the horizon $\varphi_H$ is plotted in Fig. \ref{fig:phiH_t_nonrot}. Four different black holes are presented having two different black hole masses and using both coupling functions \eqref{eq:coupling12} and \eqref{eq:coupling21} \footnote{Note that in the nonrotating case scalarization is possible only for $\epsilon>0$}. On the $x$-axis the normalized tortoise coordinate $x/\lambda$ is plotted. Two of the solutions are relatively close to  the bifurcation point having mass $M/\lambda=0.55$ \textit{(right panels)} and the other two are away from the bifurcation with $M/\lambda=0.4$ \textit{(left panels)}. The exact location of the bifurcation point in the static black hole case is at $(M/\lambda)_{\rm crit}=0.587$ and it is the same for both coupling functions. As we will show below, even for the smaller mass model the approximation we are considering, i.e. neglecting the backreaction on the metric,  gives very good results.
	
	 As expected, and also confirmed by the numerical simulations, the final black hole state does not depend on the initial data for the evolution but the path to reach this equilibrium state can be different, especially prior to the start of the exponential growth of the scalar field. In the simulations presented in Figs.   \ref{fig:Psi_evol_nonrot} and \ref{fig:phiH_t_nonrot}, the initial data is a small perturbation having a maximum close to the black hole horizon and set to zero at several back hole horizon radii. The amplitude of the perturbation is chosen to be at least two orders of magnitude smaller compared to the final state that is reached at the end of the evolution. In the case of a stable Schwarzschild black hole such perturbation would decay exponentially in time and the energy of the scalar field will be radiated away through gravitational waves. The solutions we are considering are in the unstable region, though, which means that the scalar field will be increasing exponentially until a new scalarized state is reached. This is visible in the figures: as the time advances, the initial perturbation grows exponentially and eventually it settles to a new stable state (depicted with dark red line in Fig.  \ref{fig:Psi_evol_nonrot}). The growth is slower close to the bifurcation point and the time for reaching the end state decreases for smaller black hole masses. The two coupling functions differ mainly in the strength of the final scalar field configuration -- the scalar field reaches larger values for the first coupling function, as expected also from the results in \cite{Doneva:2018rou}, but the time evolution is qualitatively very similar and the characteristic time scales are alike for a fixed black hole mass. For example, the evolution of $\varphi_H$ follows very similar trajectory in the case of coupling \eqref{eq:coupling12} and \eqref{eq:coupling21} until the scalar field backreaction starts to dominate and the field saturates to a fixed value.

\begin{figure}
	 	\includegraphics[width=0.60\textwidth]{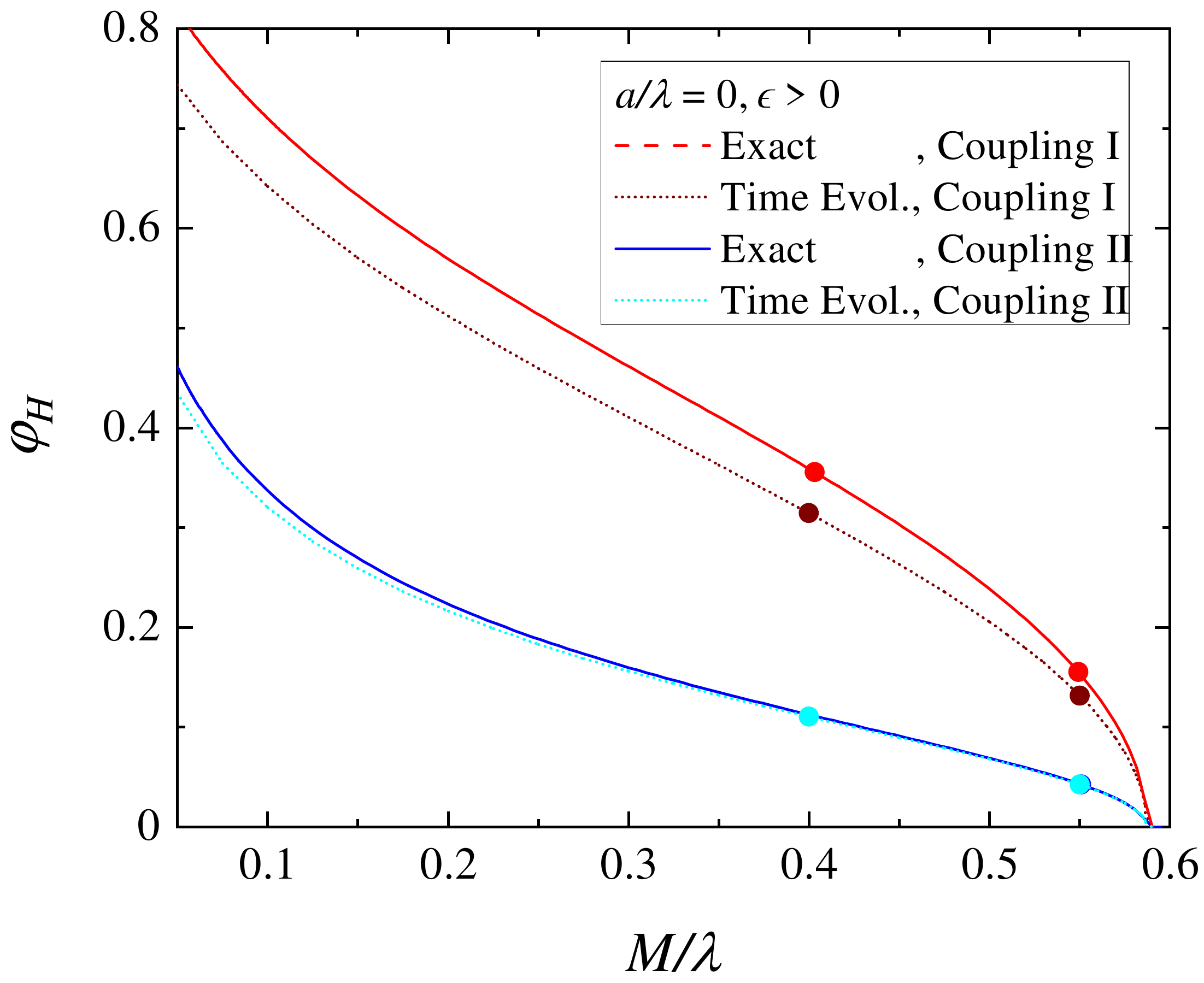}
	 	\caption{The scalar field at the horizon as a function of the black hole mass for sequences of scalarized black holes with the two coupling functions   \eqref{eq:coupling12} and \eqref{eq:coupling21}, where $\beta=12$. The solid lines correspond to the black hole models calculated in a fully consistent way using the reduced static field equations, while the dotted lines are the approximate solutions obtained after performing the time evolution of equation \eqref{eq:PertEq} on the Schwarzschild background. The particular black hole solutions presented in Figs.  \ref{fig:Psi_evol_nonrot}, \ref{fig:phiH_t_nonrot}, and \ref{fig:psi_r_merged} are denoted with dots on the corresponding sequence.}
	 	\label{fig:phiH_M_nonrot}
	 \end{figure}

	 As we have demonstrated, $\varphi_H$ tends to a constant for large enough times and the radial profile settles to an equilibrium solution. In order to check how close the obtained solutions are to the exact numerical ones, in Fig. \ref{fig:phiH_M_nonrot} we have plotted $\varphi_H$ as a function of the normalized black hole mass $M/\lambda$ for sequences of solutions obtained by solving eq. \eqref{eq:PertEq} and by solving the static nonlinear field equations \cite{Doneva:2017bvd, Doneva:2018rou}, for the two coupling functions with $\beta=12$. As one can see, close to the bifurcation point the approximate branch coming from the time evolution is very close to the true static solutions, while larger differences can be observed with the decrease of the black hole mass. Still, the differences between the two lines is small for the considered cases. Interestingly, for the second coupling function \eqref{eq:coupling21} the two lines are almost indistinguishable and that is why we will employ it in the subsequent analysis of the rotating case. As a matter of fact this is an expected result, since the studies in \cite{Doneva:2018rou} showed that even though the coupling function \eqref{eq:coupling21} produces non-negligible scalar field close to the black hole horizon as well as moderate scalar charge, the change in the metric function is almost indistinguishable from GR.
	 
	 \begin{figure}
	 	\includegraphics[width=1\textwidth]{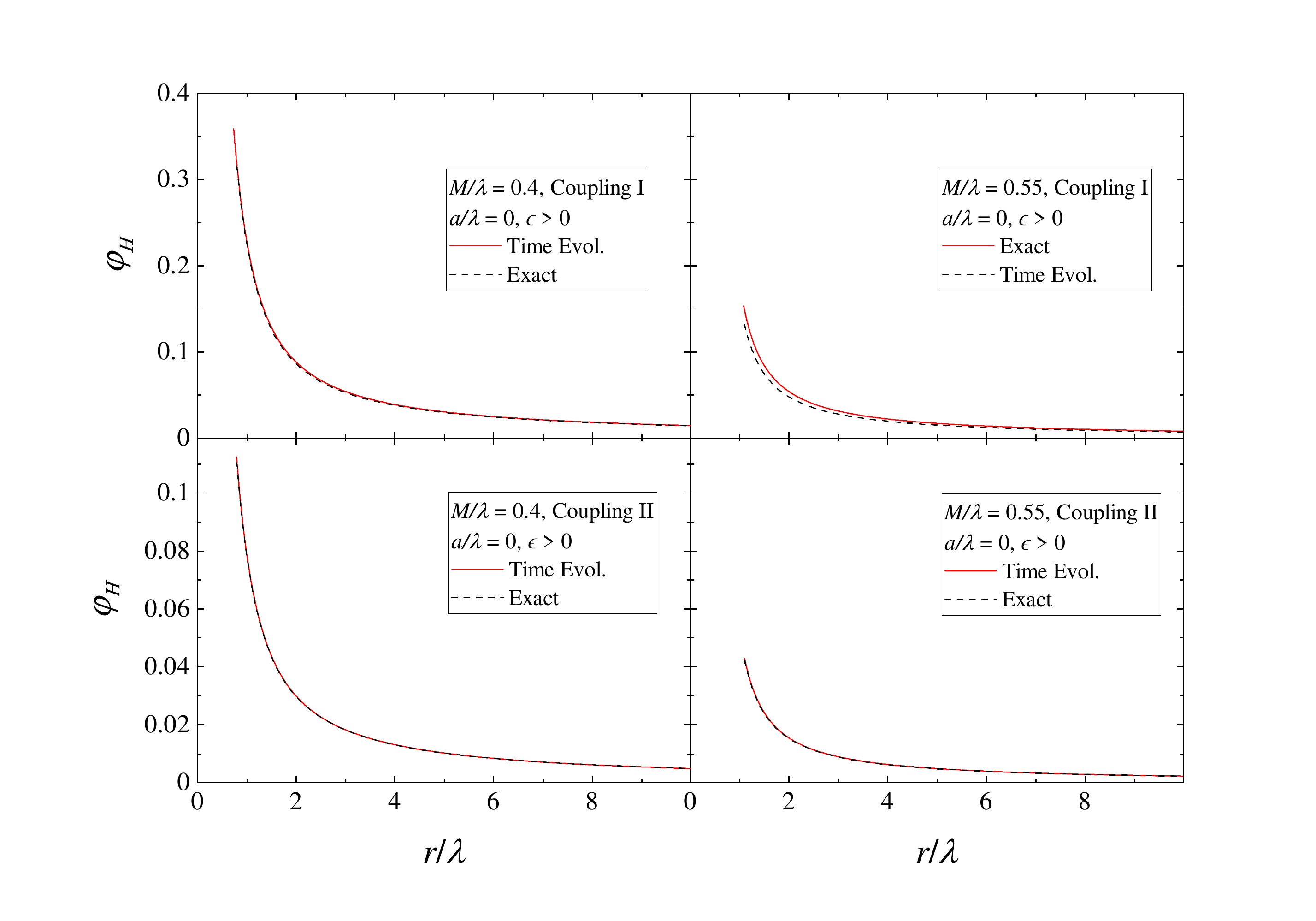}
	 	\caption{A comparison between the scalar field radial profiles  (presented in the normalized radial coordinate $r/\lambda$ on the $x$-axis) resulting from solving the full nonlinear system of reduced field equations (solid red lines) and after performing the time evolution (dashed black lines). The set of models is the same as the one presented in Figs. \ref{fig:Psi_evol_nonrot} and \ref{fig:phiH_t_nonrot}.}
	 	\label{fig:psi_r_merged}
	 \end{figure}

	 Not only the scalar field at the horizon is similar for the black holes obtained from time evolution and from the static nonlinear field equations, but as demonstrated in Fig. \ref{fig:psi_r_merged}, the whole scalar field profiles also have a small difference even away from the bifurcation point. Another important characteristic that the time evolution models should reproduce is the scalar field asymptotic and more precisely the scalar charge $D$. Extracting this quantity is not as straightforward as $\varphi_H$ since it requires a long evolution time and good resolution of the simulations. We have extracted the scalar charge for the models presented in Fig. \ref{fig:psi_r_merged} that showed a very good agreement between the two approaches -- the maximum relative difference is similar or even a bit smaller compared to the one observed for $\varphi_H$. For example, the models with the first coupling function \eqref{eq:coupling12} in Fig. \ref{fig:psi_r_merged} lead to roughly 10\% error in the determination of $D$, while for the second coupling function \eqref{eq:coupling21} the deviation is  smaller, of the order of a few percents.
	 
	At the end, let us point out one important limitations of our approximate approach  -- there is no straightforward way to show what the domain of existence of the scalarized solutions is. As we discussed, the point of bifurcation can be easily determined even if the linearized version of eq. \eqref{eq:PertEq} is considered. Apart from that a common feature for the Gauss-Bonnet black holes is that there exist a minimum mass below which the field equations do not have a solution that satisfies the regularity condition on the black hole horizon (see e.g. \cite{Doneva:2017bvd}). If one solves the static nonlinear field equations, this disappearance of the solutions for small masses is naturally obtained. We are considering here only the scalar field equation \eqref{eq:PertEq}, though, neglecting the backreaction on the metric. Thus, determining the point where the scalarized solutions cease to exists due to a violation of the regularity condition at the horizon is not directly possible. That is why one has to either derive a prior the domain of existence using the static nonlinear field equations, or stay very close to the bifurcation point. The latter option is very computationally demanding since the timescale for the development of scalarization increases with the increase of the black hole mass, and in  the limit when the bifurcation  point is reached it tends to infinity. That is why in our simulations we  have used as a reference the domain of existence presented in \cite{Doneva:2018rou} for the static case and in \cite{Cunha:2019dwb,Herdeiro:2020wei} for the rotating case and presented solutions that is inside this domain.

	\subsection{Rotating black holes}
	While for static black holes scalarization is possible only for $\epsilon>0$, this is not the case when rapid rotation is taken into account. Thus the Kerr black hole scalarize both for $\epsilon>0$ and $\epsilon<0$. The latter case happens only for rapidly rotating solutions and is called spin-induced scalarization. Below we will focus on these two cases separately since the domain of existence of solutions as well as the behavior of the equilibrium solutions is quite different. In this subsection we will work only with the coupling function \eqref{eq:coupling21} with $\beta=12$ since we have shown above that in this case the considered approximations is fulfilled with a very good accuracy. We expect that the results will be qualitatively similar for the other coupling function. A small drawback in this approach is that for this coupling no rotating black holes were obtained up to date and the domain of existence of solutions can not be determined with a good accuracy. Still, the analysis in the nonrotating case shows, that the static scalarized black holes exist at least for the same range of masses as the coupling \eqref{eq:coupling12} with $\beta=6$ (the one considered in \cite{Cunha:2019dwb,Herdeiro:2020wei}). Moreover, a general observation in \cite{Doneva:2018rou} is that the increase of $\beta$ leads to a larger domain of existence of the solutions while the deviations from GR are suppressed. That is why  our choice of coupling function and $\beta$ is justified and very suitable for the employed approximation.
	 
	\subsubsection{Standard case when $\epsilon>0$.}
	Scalarized rotating black holes were first obtained in this case \cite{Cunha:2019dwb} since they have a natural nonrotating limit $a \rightarrow 0$. We have focused on two models with $M/\lambda=0.56$, $a/\lambda=0.3$ and  $M/\lambda=0.59$, $a/\lambda=0.5$. Both of them are chosen to be very close to the bifurcation point so that the scalar field is small and the backreaction of the scalar field on the spacetime geometry can be safely neglected.	In addition, in these simulations we have used different initial data compared to the static case just for the sake of completeness and to be able to demonstrate the different paths in reaching an equilibrium solution. Thus we start the simulations with a Gauss pulse (in radial direction) located far outside the black hole horizon and having initial amplitude around two orders of magnitude smaller than the expected maximum of $\varphi_H$ of the equilibrium solution.
	
\begin{figure}
		\includegraphics[width=0.45\textwidth]{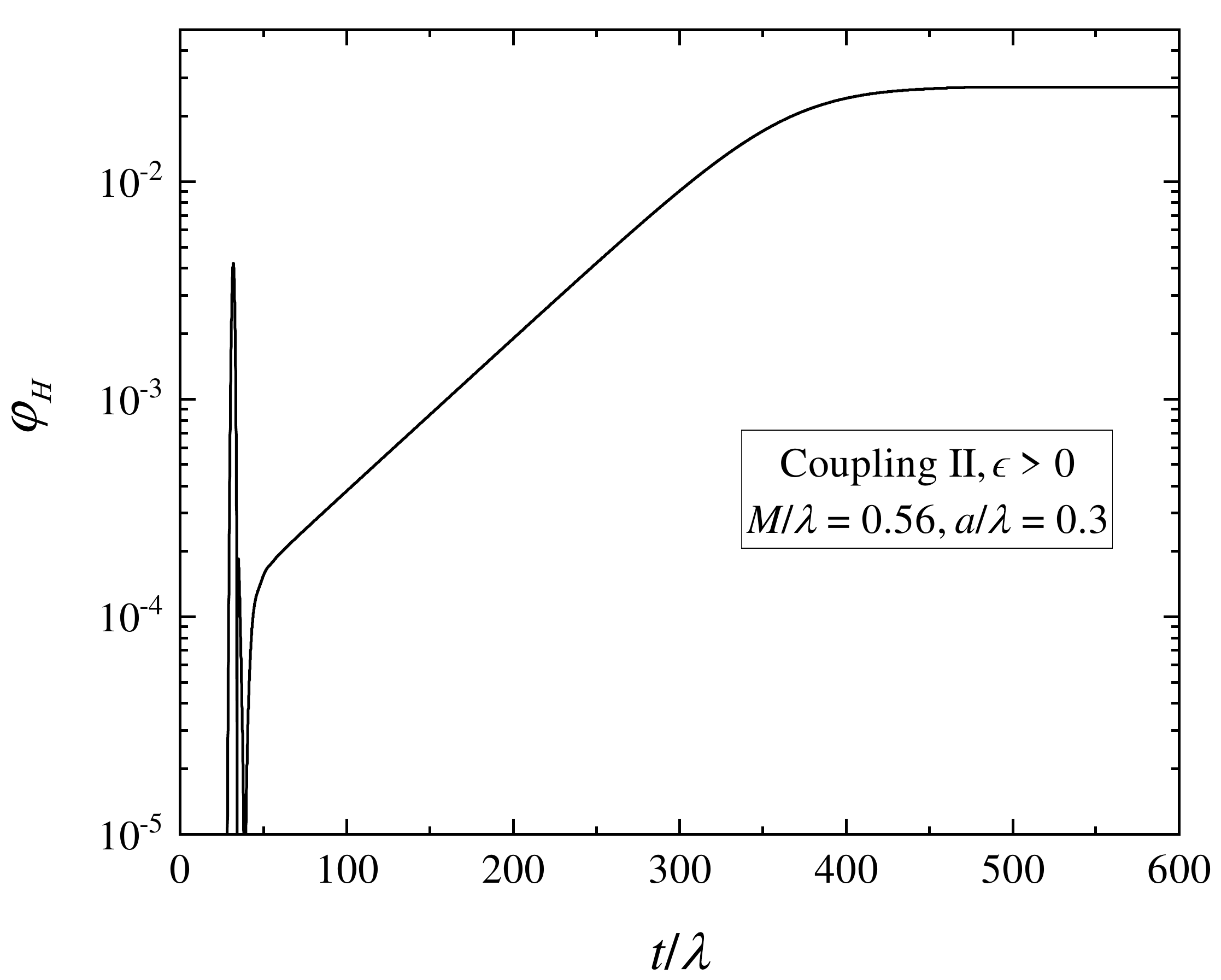}
		\includegraphics[width=0.45\textwidth]{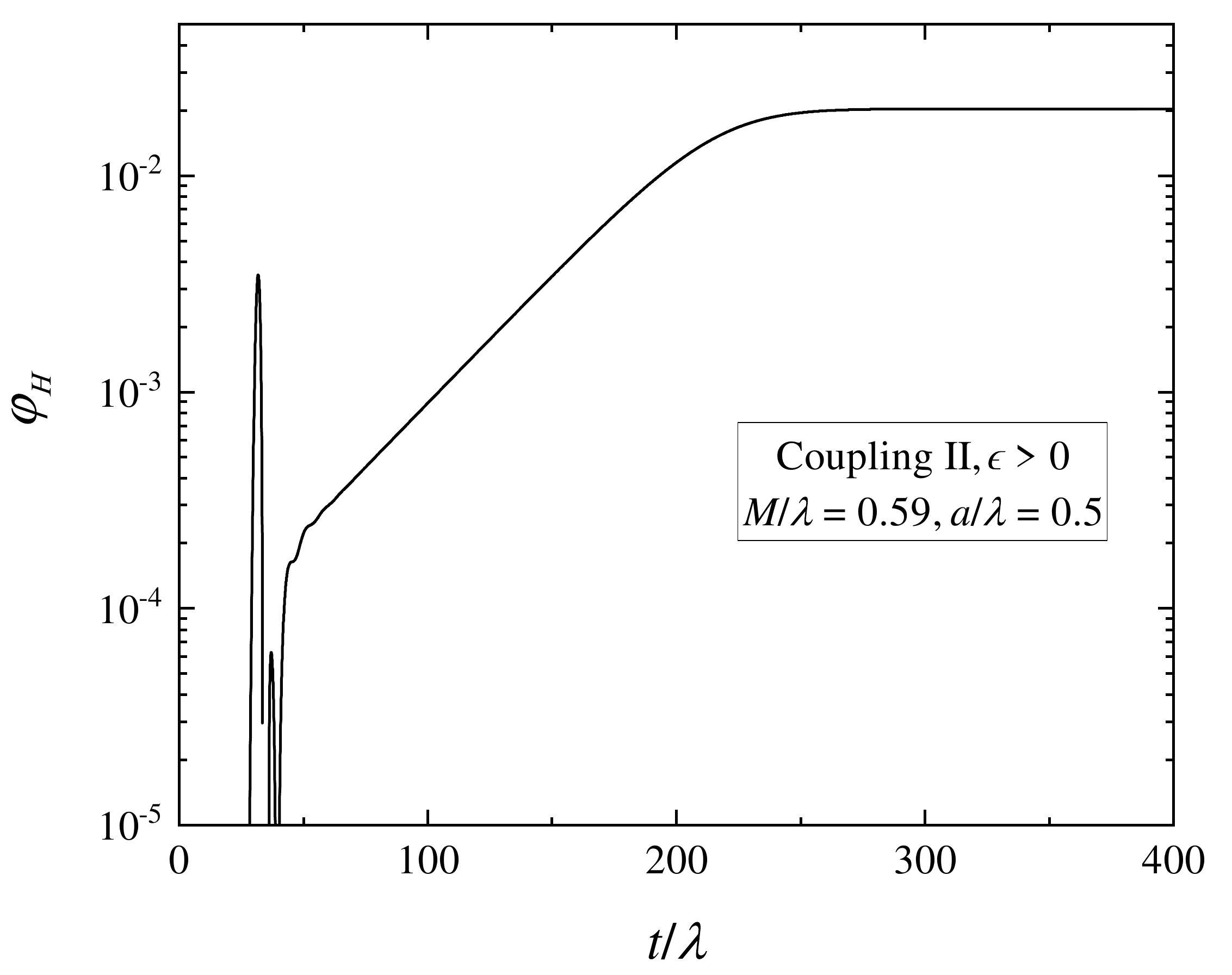}
		\caption{The  scalar field at the horizon, extracted at $\theta=\pi/4$, as a function of time for two rotating black holes with $\epsilon>0$: a moderately rotating model with $M/\lambda=0.56$, $a/\lambda=0.3$ \textit{(left panel)} and a more rapidly rotating model with $M/\lambda=0.59$, $a/\lambda=0.5$ \textit{(right panel)}. The results are for the second coupling function \eqref{eq:coupling21} with $\beta=12$.}
		\label{fig:psi_t_M03_a024_21_beta12_lamb-1}
	\end{figure}

\begin{figure}
		\includegraphics[width=0.45\textwidth]{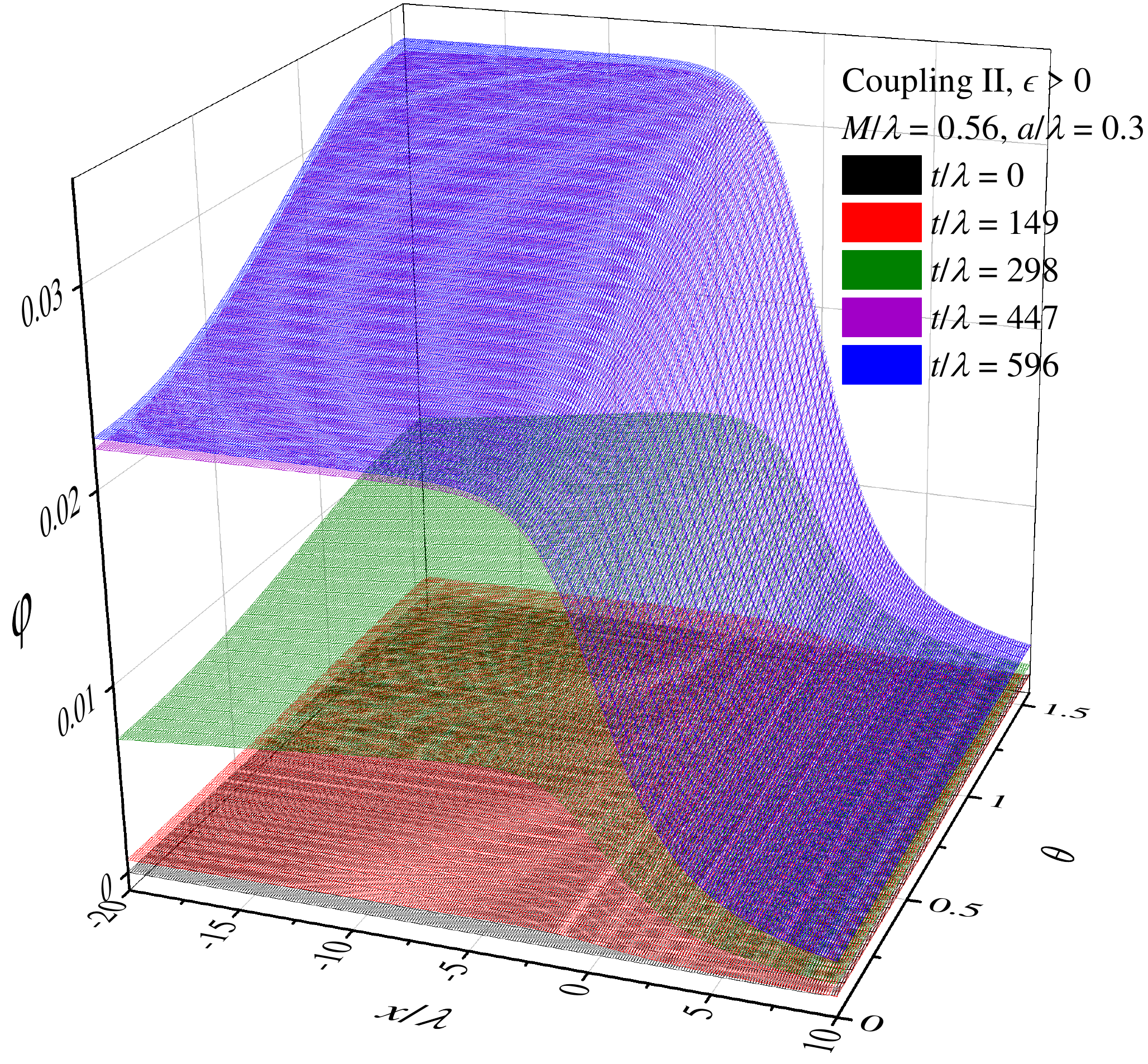}
		\includegraphics[width=0.45\textwidth]{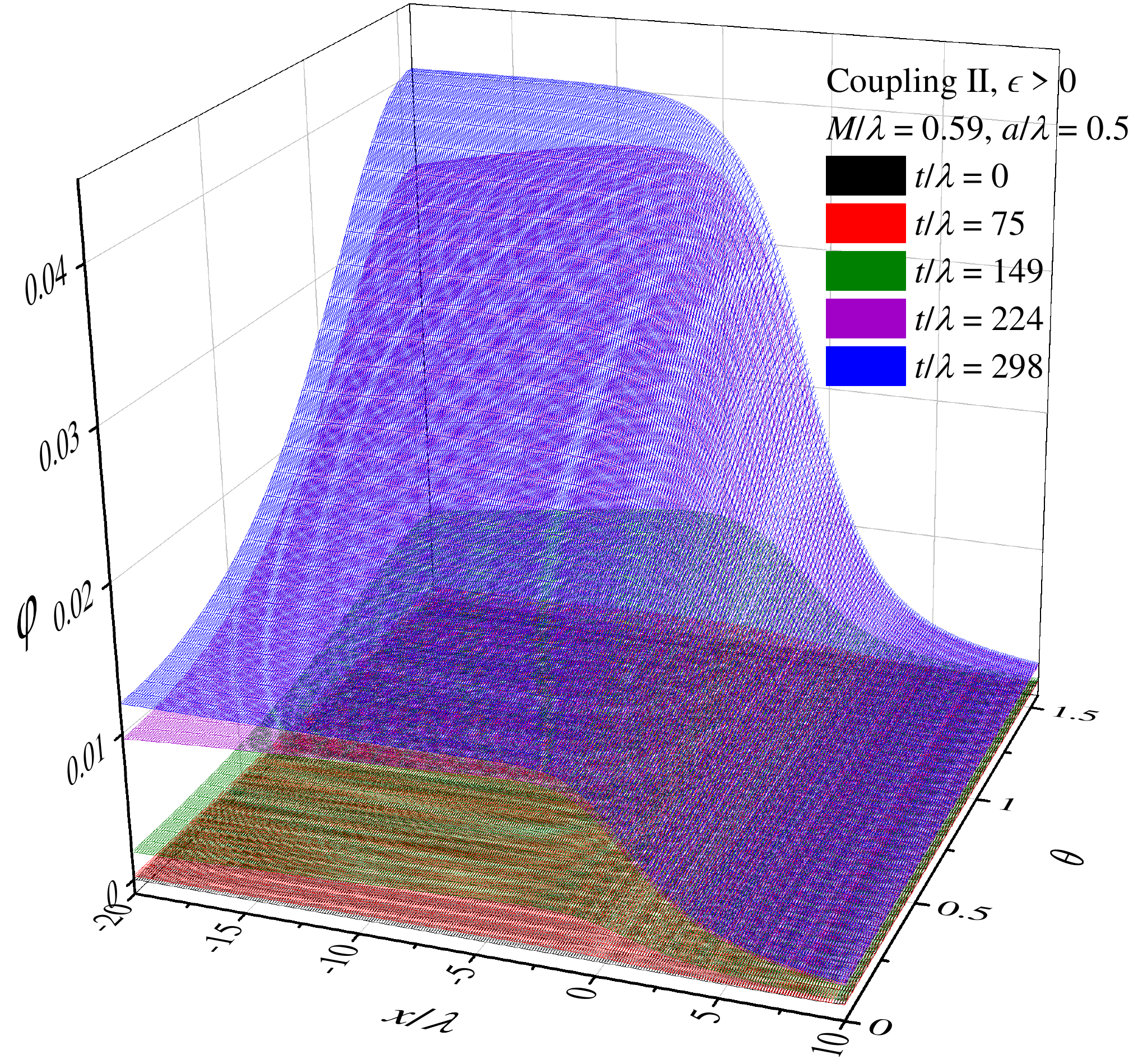}
		\caption{Snapshots of the scalar field profile during the evolution. The models are the same as the ones presented in Fig. \ref{fig:psi_t_M03_a024_21_beta12_lamb-1}.}
		\label{fig:3D_M03_a024_12_beta12_lamb-1}
	\end{figure}

	The time evolution of the scalar field of the horizon taken at $\theta=\pi/4$ is plotted in Fig. \ref{fig:psi_t_M03_a024_21_beta12_lamb-1} while snapshots of the scalar field profile at different times is depicted in Fig. \ref{fig:3D_M03_a024_12_beta12_lamb-1}. One can notice that for such initial data first a few damped oscillations are observed, since the incoming pulse excites the black hole scalar modes. The instability develops quickly and these oscillations are followed by an exponential growth of the scalar field and a subsequent saturation due to the scalar field backreaction. The timescale of the scalarization is larger compared to the presented static black holes, because we have considered models that are very close to the bifurcation point. The two-dimensional profile of the scalar field when equilibrium is reached shows a minimum at $\theta=0$ and a maximum at $\theta=\pi/2$. As expected, the $\theta$ derivative of the scalar field at this points tends to zero.    
	
	\subsubsection{Spin induced scalarization with $\epsilon<0$}
	In the case of spin induced scalarization we have considered two representative cases  $M/\lambda=0.3$, $a/\lambda=0.24$ and $M/\lambda=0.8$, $a/\lambda=0.79$. Here scalarization is possible only for rapidly rotating black holes and we have chosen both models more or less close to the extremal limit. The time evolution of the scalar field on the horizon taken at $\theta=\pi/4$ is plotted in Fig. \ref{fig:psi_t_M03_a024_21_beta12_lamb-1} while snapshots of the scalar field profile at different times is depicted in Fig. \ref{fig:3D_M03_a024_12_beta12_lamb-1}. The initial data are the same as the ones used in the previous subsection with $\epsilon>0$. The qualitative behavior of the scalar field evolution is quite similar to the one observed for $\epsilon>0$ and the main difference comes from the change of the equilibrium black hole profile reached at later times. Contrary to the previous case, here the scalar field forms a maximum at the $\theta=0$ and a minimum at $\theta=\pi/2$ with a $\theta$ derivative vanishing at these points. This shows, that the geometry of the scalar field is quite different for scalarization with $\epsilon<0$ and $\epsilon>0$. This can potentially have some  observational manifestations and provide ways to distinguish between both of them.

	\begin{figure}
		\includegraphics[width=0.45\textwidth]{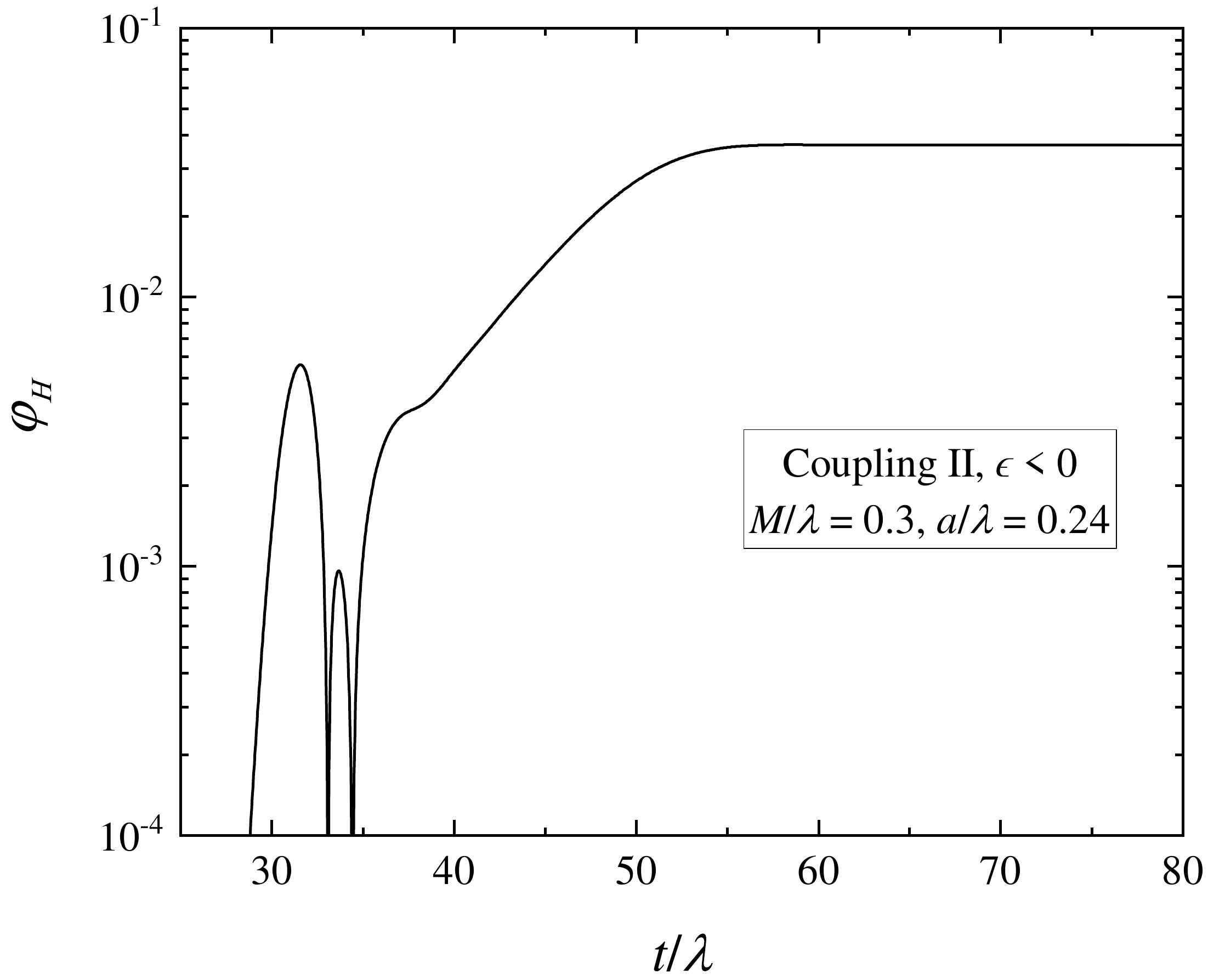}
		\includegraphics[width=0.45\textwidth]{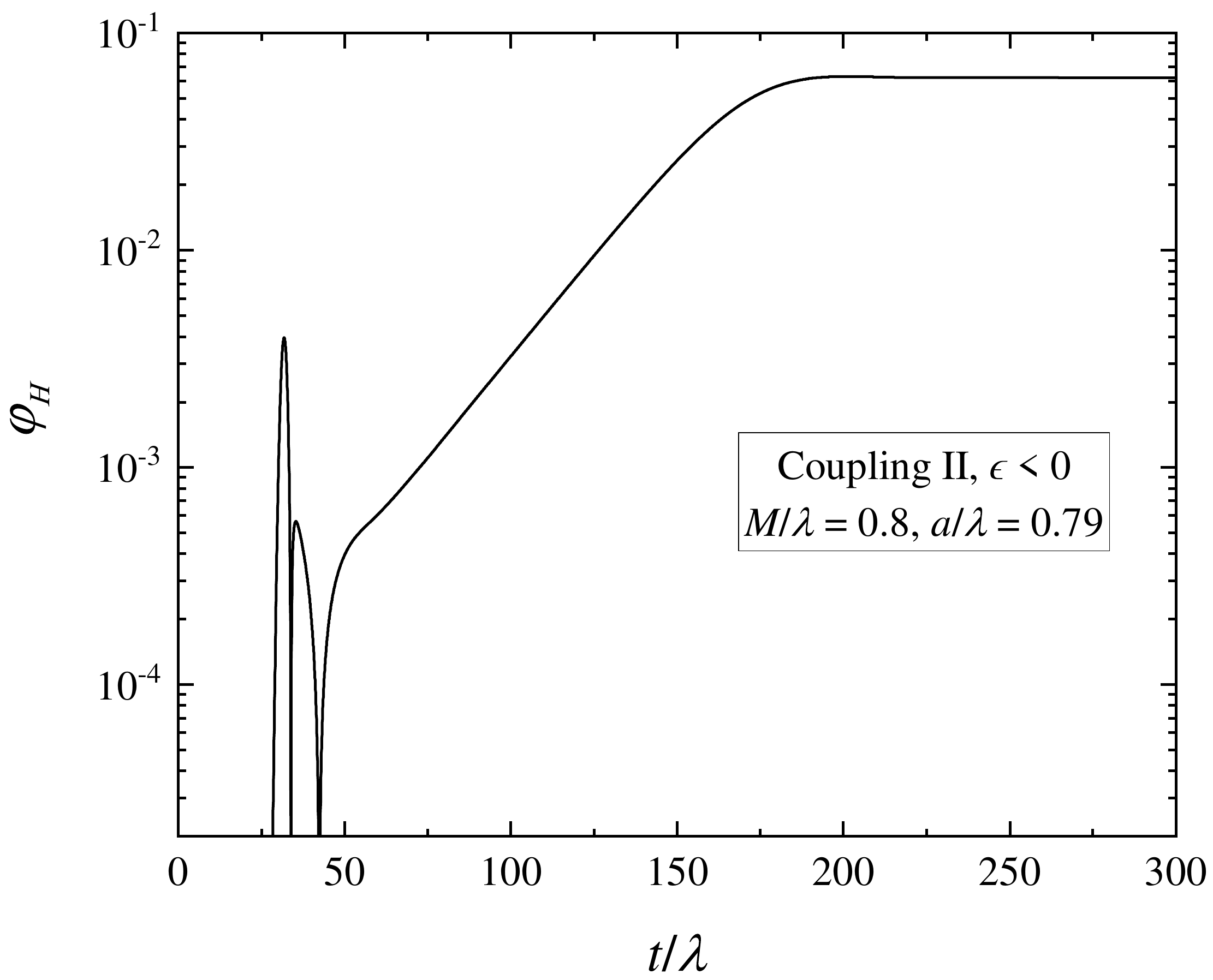}
		\caption{The  scalar field at the horizon, extracted at $\theta=\pi/4$, as a function of time for rotating black holes with $\epsilon<0$ (spin-induced scalarization): a more moderately rotating model with $M/\lambda=0.3$, $a/\lambda=0.24$ \textit{(left panel)} and a model close to the extremal limit with$M/\lambda=0.8$, $a/\lambda=0.79$ \textit{(right panel)}. The results are for the second coupling function \eqref{eq:coupling21} with $\beta=12$.}
		\label{fig:psi_t_M03_a024_21_beta12_lamb1}
	\end{figure}
	
	\begin{figure}
		\includegraphics[width=0.45\textwidth]{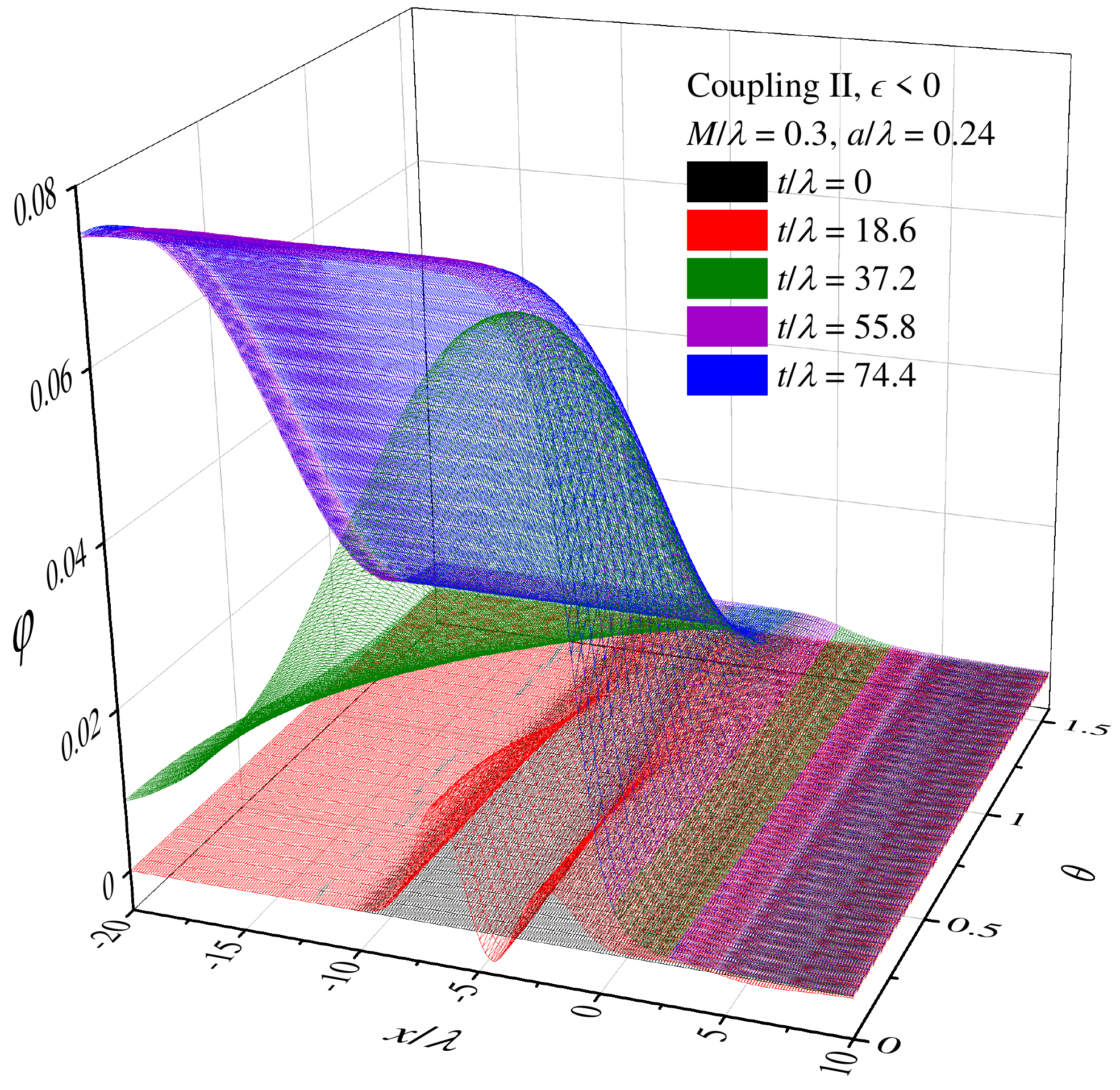}
		\includegraphics[width=0.45\textwidth]{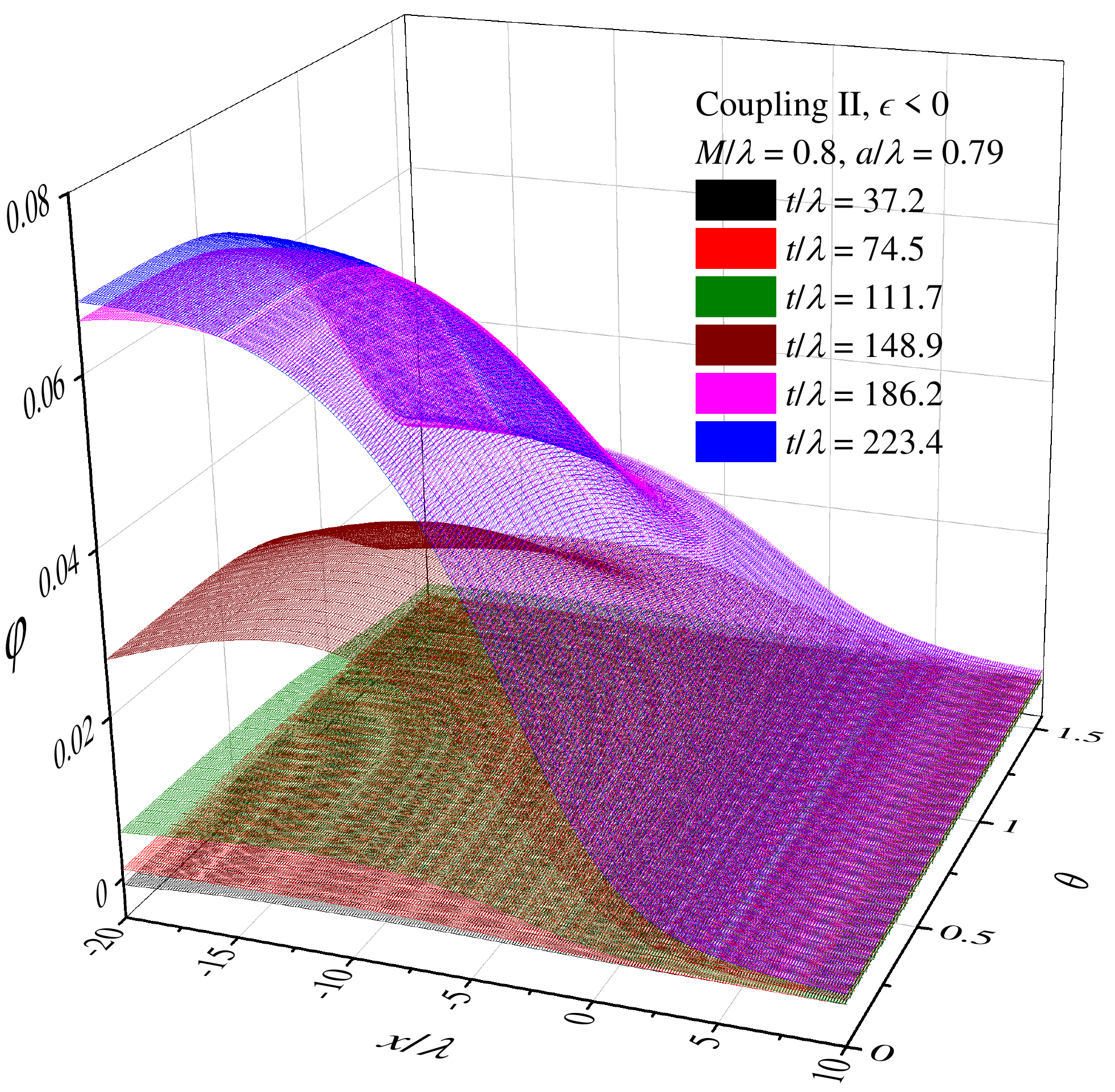}
		\caption{Snapshots of the scalar field profile during the evolution. The models are the same as the ones presented in Fig. \ref{fig:psi_t_M03_a024_21_beta12_lamb1}.}
		\label{fig:3D_M03_a024_12_beta12_lamb1}
	\end{figure}
	
	\section{Conclusion}
	In the present paper we have studied the dynamical development of black hole scalarization in Gauss-Bonnet gravity. We have focused both on static and rotating solutions. The calculations are performed in the  decoupling limit, i.e. the nonlinear scalar field equation is evolved on the Schwarzschild and Kerr background starting from a small perturbation. In the region where the Kerr black hole loses stability this perturbation will grow exponentially until it settles to a stationary  equilibrium state.  The simulations confirm that while the path to reach this equilibrium black hole might be strongly dependent on the initial perturbation, the end state is unaltered. In our simulations we have focused both on static and rotating scalarized black holes.
	
	The Schwarzschild black hole can scalarize only when the second derivative of the coupling function at $\varphi=0$ is greater than zero, i.e. $\frac{d^2 f(0)}{d\varphi^2}  > 0$. Kerr black holes can develop scalar hair both for $\frac{d^2 f(0)}{d\varphi^2}  > 0$ and $\frac{d^2 f(0)}{d\varphi^2}  < 0$. The latter option can realize only for rapid rotation and is called spin-induced scalarization. In the present paper we have considered all of these cases.
	
	The static case was examined in greater detail in order to evaluate the accuracy of the decoupling approximation. As the bifurcation point is approached this assumption leads to more and more accurate results as the magnitude of the scalar field decreases. Our goal, though, was to evaluate how accurate this description is also away from the bifurcation. For this purpose several coupling functions have been considered. Using as a reference the static solutions obtained in \cite{Doneva:2018rou}, we have compared different characteristics such as the scalar charge, the scalar field at the horizon, as well as the whole scalar field profile for certain representative models. It turns out that for one of the considered coupling functions the match between the time evolution equilibrium solution and the one obtained from the static nonlinear field equations is almost in a perfect agreement. For the rest of the couplings still in a very good agreement is observed. Thus we have shown explicitly, that if we limit ourselves to solutions with relatively low scalar charge, the considered decoupling approximation gives good results even away from the bifurcation point.
	
	With this motivation in hand, we have evolved the scalar field on the Kerr background and observed the development of scalarization. In the cases of $\frac{d^2 f(0)}{d\varphi^2}  > 0$ and $\frac{d^2 f(0)}{d\varphi^2}  < 0$ the equilibrium scalar field profile has quite different characteristics -- while in the former case the scalar field has a minimum at the rotational axes and maximum at $\theta=\pi/2$ (for a constant radial coordinate), this is reversed for the spin induced scalarization where the scalar field peaks at the axes.
	
	\section*{Acknowledgements}
	DD acknowledges financial support via an Emmy Noether Research Group funded by the German Research Foundation (DFG) under grant
	no. DO 1771/1-1.  SY would like to thank the University of T\"ubingen for the financial support.  
	The partial support by the Bulgarian NSF Grant KP-06-H28/7 and the  Networking support by the COST Actions  CA16104 and CA16214 are also gratefully acknowledged. 
	
	
    \bibliographystyle{apsrev4-2}
	 \bibliography{references}

\end{document}